\documentclass[9pt,twocolumn,twoside]{osajnl}
\journal{ol} % Choose journal (ao, aop, josaa, josab, ol, optica, pr)

\setboolean{shortarticle}{true}
\usepackage{upgreek}
\usepackage{units}

\title{High-quality Tailored-edge Cleaving Using Aberration-corrected Bessel-like Beams}

\author[1,2,6*]{Michael Jenne}
\author[1,6]{Daniel Flamm}
\author[1]{Taoufiq Ouaj}
\author[3]{Julian Hellstern}
\author[1]{Jonas Kleiner}
\author[1,4]{Daniel Großmann}
\author[1]{Maximilian Koschig}
\author[1]{Myriam Kaiser}
\author[1]{Malte Kumkar}
\author[2,5]{Stefan Nolte}

\affil[1]{TRUMPF Laser- und Systemtechnik GmbH, Johann-Maus-Strasse 2, 71254 Ditzingen, Germany}
\affil[2]{Institute of Applied Physics, Abbe Center of Photonics, Friedrich Schiller University Jena, Albert-Einstein-Strasse 15, 07745 Jena, Germany}
\affil[3]{TRUMPF Laser GmbH, Aichhalder Str. 39, 78713 Schramberg, Germany}
\affil[4]{Chair for Laser Technology, RWTH Aachen University, Steinbachstr 15, 52074 Aachen, Germany}
\affil[5]{Fraunhofer Institute for Applied Optics and Precision Engineering, Albert-Einstein-Strasse 7, 07745 Jena, Germany}
\affil[6]{Both authors contributed equally}

\affil[*]{Corresponding author: michael.jenne@trumpf.com}

 \dates{Compiled May 8, 2018}

\ociscodes{(320.0320) Ultrafast optics; (070.7145) Ultrafast processing; (100.0118) Imaging ultrafast phenomena, (140.7090) Ultrafast lasers; (160.2750) Glass and other amorphous materials; (120.2040) Displays; (140.3390) Laser materials processing.}

\doi{\url{https://doi.org/10.1364/OL.43.003164}\\ \\ \copyright \,\, \textbf{2020 Optical Society of America. One print or electronic copy may be made for personal use only. Systematic reproduction and distribution, duplication of any material in this paper for a fee or for commercial purposes, or modifications of the content of this paper are prohibited.\\}}

\begin{abstract}
We report on the usage of ultrashort laser pulses in form of aberration-corrected Bessel-like beams for laser cutting of glass with bevels. Our approach foresees to incline the material's entrance surface with respect to the processing optics. The detailed analysis of phase distortions caused by the beam transition through the tilted glass surface allows to pre-compensate occurring aberrations using digital holography. We verify theoretical considerations by means of pump-probe microscopy and present high-quality edges in non-strengthened silicate glass.
\end{abstract}

\setboolean{displaycopyright}{true}

\begin{document}

\maketitle

\noindent Throughout recent years ultrashort laser pulses have experienced great interest in research and industry, which led to a broad range of applications ranging from spectroscopic applications to fabrication of 3D objects \cite{Malinauskas2016, Luo06}. Especially the possibility to interact with wide band-gap materials enables the precise processing of, e.g. semi-conductors used in thin-film technologies, or technical glasses, commonly used for displays and safety glasses and is of great interest for markets like consumer electronics \citep{Gattass:08, Kumkar14}. 
Sensitive tuning of the high intensities that trigger the initial nonlinear interaction processes and the high temporal resolution due to the short interaction times determines if the interaction will lead to ablation, welding or cleaving \cite{Ito2018, Watanabe2016}. The use of lasers for glass cutting is becoming more and more attractive compared to conventional methods such as waterjet cutting, as process speed and quality constantly improve \cite{Nisar13} and the possibility of beam shaping allows access to new cutting strategies such as curved edges \cite{Mathis2012}. Bessel-like beams are promising candidates because they have an elongated but sharp profile, which enables single pass cleaving of glasses with ticknesses up to several milimeters. To increase crack length and crack orientation for improved throughput and minimized damage current studies investigate glass processing with elliptical or aberrated Bessel-like beams \cite{Meyer17, Dudutis18}. 
However, new products on the smartphone market represent the trend to new and arbitrary display shapes with tailored edges detached from the rigid right-angled predecessors. 
\begin{figure}[]
\centering
\includegraphics[width=\linewidth]{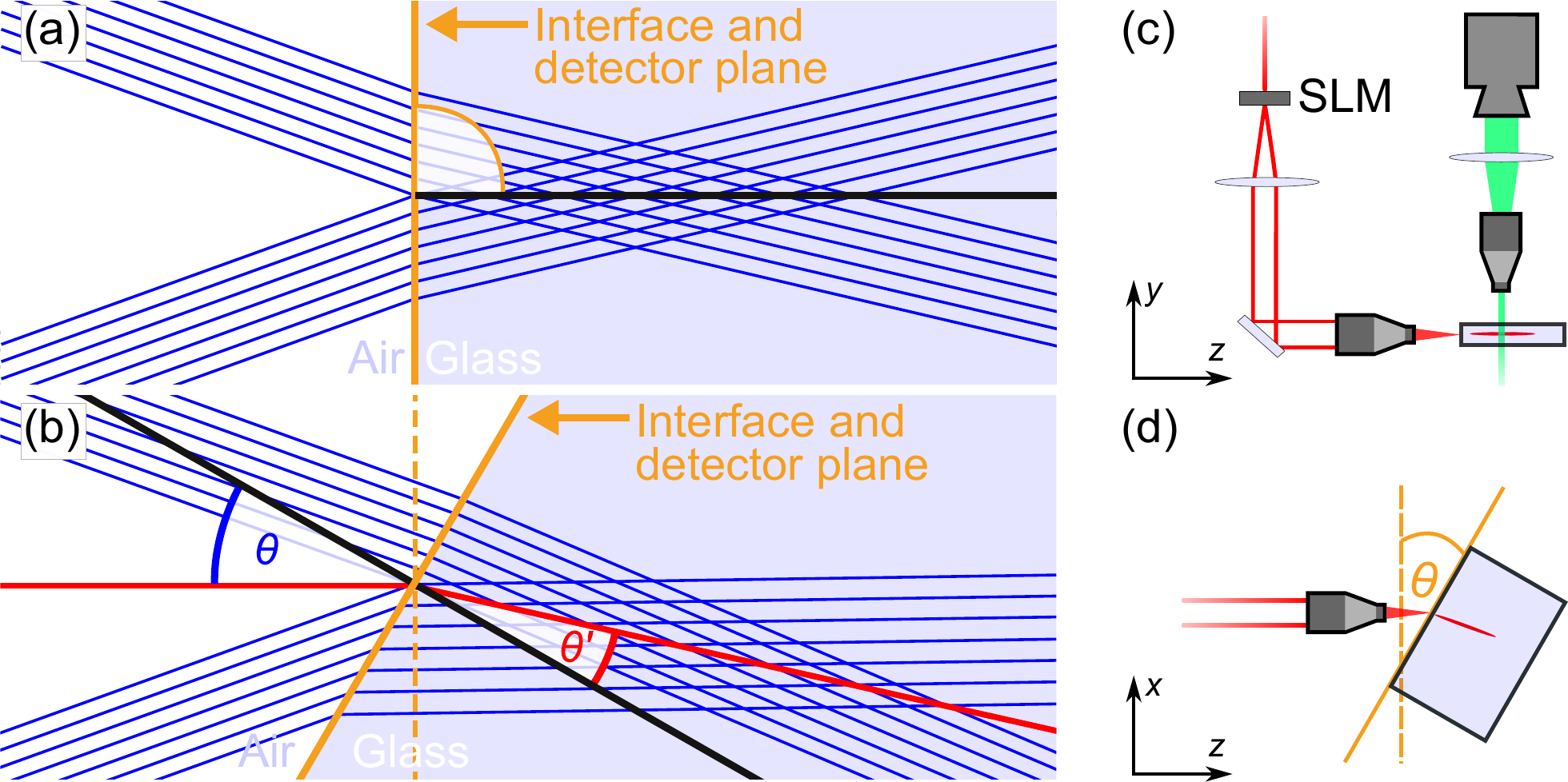}
\caption{Geometric-optical representation of the Bessel-like focus situation behind the glass surface for incident illumination $\theta = 0$ (a) and for inclined illumination $\theta > 0$ (b). Corresponding schematic of the transverse pump-probe microscope (c). Pump pulses at $\unit[1030]{nm}$ and $\unit[900]{fs}$ pulse duration are beam-shaped by the SLM and focused into the sample using a $4f$-setup. The resulting light-matter interaction is probed by a $\unit[515]{nm}$ laser pulse with $\unit[200]{fs}$ pulse duration. Rotatable mounted glass with definition of cleaving angle $\theta$ (d).}
\label{fig:Figure1}
\vspace{-0.5cm}
\end{figure}
To generate beveled edges, which can be regarded as the simplest form of advanced edge-profiles, the following work analyzes the aberrations of an one axis interface tilt to the performance attributes of classical cleaving processes with Bessel-like beam profiles. Our approach is based on the correction of occurring aberrations, in-situ pump-probe microscopy thereof and finally the successful implementation into a cleaving process.

\noindent Generation of Bessel-like beams is often axicon-based. In our case, a ``digital'' axicon \cite{Trichili2014, Bergner2018} is used and displayed by a liquid-crystal-on-silicon-based spatial light modulator (SLM). Within the approximation of thin elements, a real axicon with given angle $\alpha$ and refractive index $n$ can be described by a radially symmetric, phase-only transmission $T_{\text{ax}}\left(r\right) = \exp{\left(\imath \beta r\right)}$ with linear slope value $\beta$ determined by $\beta = 2\uppi\left(n-1\right)\alpha / \lambda$ \cite{Leach2006, Bergner2018}. A $4f$-setup with 20\,$\times$\,telescopic demagnification scales the shaped optical field of the SLM plane to the required micrometer dimensions and focuses it into the glass volume \cite{Flamm2015}. Using this imaging concept, the SLM is placed virtually into the plane of the tilted glass surface. Usually, when light is focused deep into a transparent material with a plane surface spherical aberrations occur \cite{Itoh2009}. Bessel-like beams, on the other hand, exhibit a natural resistance to those aberrations -- a further outstanding property and an additional reason for their frequent usage for glass-cutting applications. At the interface between the two media, all radial field components exhibit the same absolute angle of incidence value and propagate identical optical path lengths until they interfere constructively to the desired elongated focus, see geometric-optical representation of an incident Bessel beam (profile cut) in \hyperref[fig:Figure1]{Fig.\,\ref{fig:Figure1}(a)}. Considering the scenario of a tilted glass surface as depicted in \hyperref[fig:Figure1]{(b)}, the spatially varying angles of refraction determined by Snell's law will break the radial symmetry and, depending on the tilt angle, will result in a characteristic interference pattern along the geometric focus zone (red line), see also \href{https://osapublishing.figshare.com/articles/Visualization1_mp4/6081140}{Visualization\,1}. This is accompanied by a tremendous loss of peak intensity and focal depth.

\begin{figure}[t]
	\centering
	\includegraphics[width=0.95\linewidth]{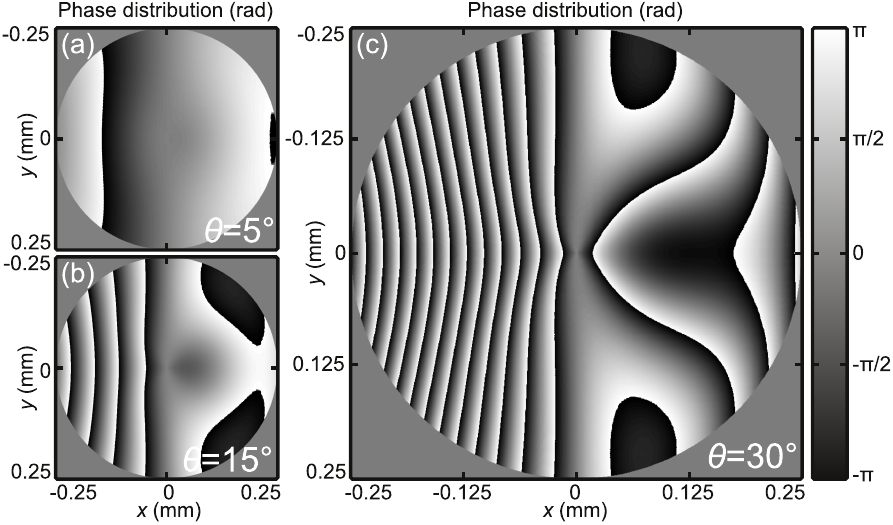}
	\caption{Simulated phase aberrations $\Delta\phi\left(x,y\right)$ caused by the beam transition through the glass interface for tilt angles of $\theta = \unit[5]{^\circ}$ (a),  $\theta = \unit[15]{^\circ}$ (b) and  $\theta = 30^\circ$ (c).}
	\label{fig:Figure2}
	\vspace{-0.5cm}
\end{figure}

\noindent For further analyzing the problem at hand more quantitatively, we use the physical optics simulation engine of VirtualLab Fusion \cite{Wyrowski2015} providing fast access to the optical field behind the (not necessarily) plane and tilted interface even for high NA situations as used in our setup, cf. \hyperref[fig:Figure1]{Fig.\,\ref{fig:Figure1}(c)} and \hyperref[fig:Figure1]{(d)}. For this purpose, we virtually place field-detectors in the plane of the glass surface for normal incidence, see \hyperref[fig:Figure1]{Fig.\,\ref{fig:Figure1}(a)}, and in the plane perpendicular to the refraction angle for tilted incidence, see \hyperref[fig:Figure1]{(b)}, respectively. Knowledge about the aberrated optical field $E_{\text{ab}}$ in amplitude and phase in the plane of the interface $E_{\text{ab}}\left(x, y\right) = A_{\text{ab}}\left(x, y\right)\exp\left[\imath\phi_{\text{ab}}\left(x, y\right)\right]$ enables to calculate deviations from the undisturbed field $E_{\text{id}}\left(x, y\right) = A_{\text{id}}\left(x, y\right)\exp\left[\imath\phi_{\text{id}}\left(x, y\right)\right]$ generating the ideal Bessel-like beam. Due to small differences for the respective Fresnel transmission coefficients ($<\unit[5]{\%}$) we assume $A_{\text{ab}} \approx A_{\text{id}}$ and focus on deviations of the corresponding phase distributions $\Delta\phi = \phi_{\text{id}} - \phi_{\text{ab}}$. Results of these simulations are depicted in \hyperref[fig:Figure2]{Fig.\,\ref{fig:Figure2}}, where $\Delta\phi$ is plotted as an example for three different tilt angles $\theta$ defined within a circle of radius $R_{\phi}' = \unit[250]{\upmu m}$.
\begin{figure}[]
	\centering
	\includegraphics[width=\linewidth]{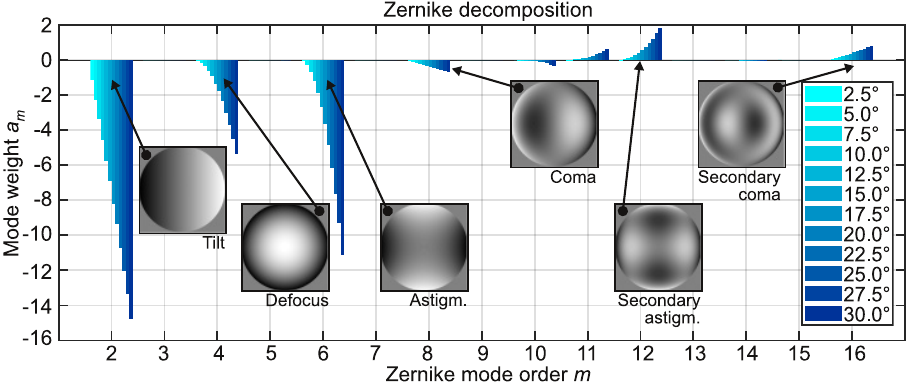}
	\caption{Decomposition of the aberrated phase profiles -- some of them depicted in \hyperref[fig:Figure2]{Fig.\,\ref{fig:Figure2}} -- into a set of $15$ Zernike modes \cite{Noll1976}. Contributing modes show a continuous growth with increasing tilt angle $\theta$ of the plane glass interface.}
	\label{fig:Figure3}
	\vspace{-0.5cm}
\end{figure}
\noindent As expected, phase aberrations are growing for increasing values of $\theta$. This is illustrated even more clearly in \hyperref[fig:Figure3]{Fig.\,\ref{fig:Figure3}} where we perform a decomposition of these phase aberrations into a set of Zernike modes $\left\{Z_{m}\left(x,y\right)\right\}$ of order $m$ (normalized and indexed according to Noll \cite{Noll1976}) using $\Delta\phi = \arg\left[\exp\left(\imath\uppi \sum_{m=1}^{m_{\text{max}}} a_mZ_m \right)\right]$ \cite{Schulze2013}. With a sufficiently large number of modes, knowledge of the associated, real-valued coefficients $a_m$ completely describe $\Delta\phi$.
\noindent The graph shows weights of coefficients $a_m$ starting from $m=2$, thus neglecting piston, up to order $m_{\text{max}}=16$ for $12$ different tilt angles between $\theta = 2.5^\circ$ and $30^\circ$. Characteristic for the problem at hand is the continuous growth of certain coefficients with increasing $\theta$ such as $m=2$ (tilt), $4$ (defocus), $6$ (oblique astigmatism) or $12$ (vertical secondary astigmatism) representing the modes that contribute most to the description of $\Delta\phi$.

\noindent As already described earlier, our processing optic design simply consists of the SLM and a telescopic $4f$-setup, cf. \hyperref[fig:Figure1]{Fig\,\ref{fig:Figure1}(c)}. This arrangement allows for both Bessel-like beam generation and compensation of occurring aberrations using a single digital hologram. For this purpose, the calculated phase aberrations $\Delta\phi\left(x,y\right)$ are inverted, the spatial scaling is adapted to the telescopic magnification of $M=20$ and the resulting phase distribution is multiplexed with the axicon-generating phase mask $T_{\text{ax}}$. Then, the final phase-only hologram $T_{\text{tot}}$ displayed by the SLM reads as $T_{\text{tot}} = T_{\text{ax}}T_{\text{ab}} = \exp{\left[\imath\left(\beta r - \Delta\phi\right)\right]}$, now defined on a magnified circle of radius $R_{\phi}=MR_{\phi}' = \unit[5]{mm}$, for our particular case. Since this phase distribution $\arg\left(T_{\text{tot}}\right)$ is dominated by the carrier grating of the digital axicon \cite{Bergner2018}, we refrain from showing the final SLM-mask at this point.

\begin{figure}[t]
	\centering	
	\includegraphics[width=1.0\linewidth]{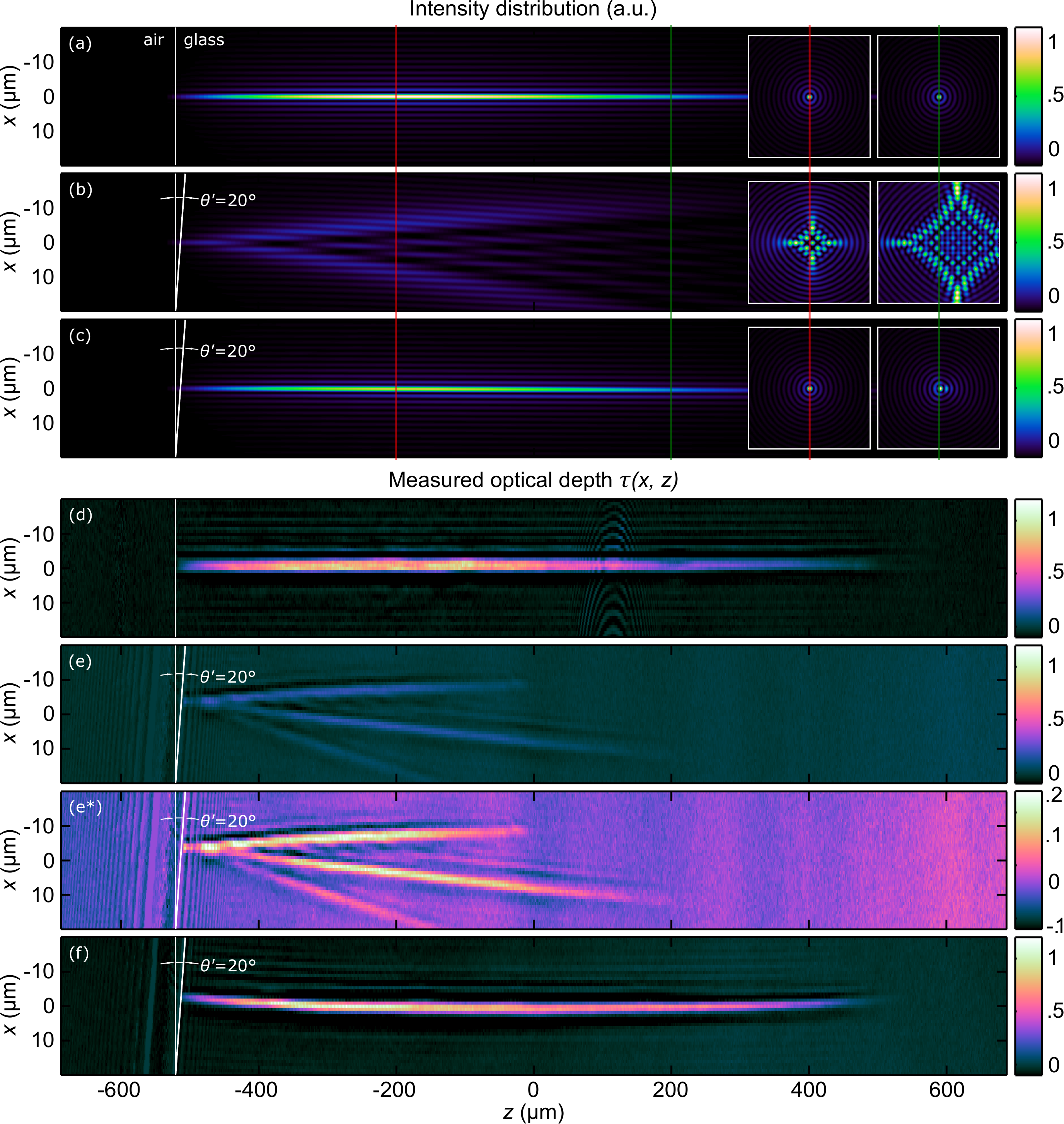}
	\caption{Simulated intensity distribution of Bessel-like beams passing the surface perpendicular (a), with tilting angle $\theta=\unit[30]{^\circ}$ (b) and after the aberration correction (c). The insets show the cross sections of the $x$-$y$-plane corresponding to (a)-(c) at $z=\unit[-200]{\upmu m}$ (red line) and $z=\unit[200]{\upmu m}$ (green line), respectively. Corresponding measured pump-probe images show the calculated optical depth for perpendicular (d) and tilted (e and with different intensity scale e*) incidence. The corrected absorption profile in (f) shows the restored features of a Bessel-like beam. Please note that the $x$-axis is stretched by a factor of 5.}
	\label{fig:Figure4}
	\vspace{-0.5cm}
\end{figure}

\noindent The efficacy of this approach is emphasized in \href{https://osapublishing.figshare.com/articles/Visualization2_mp4/6081125}{Visualisation\,2} where three caustic ``flights'' can be compared. First, the case of the non-aberrated Bessel-like focus is shown ($\theta = 0^\circ$). The second displays the aberrated one for $\theta = 30^\circ$. Finally, the aberration correction is applied for the same angle of incidence restoring the original focus distribution almost perfectly, cf. \hyperref[fig:Figure4]{ Fig.\,\ref{fig:Figure4}(a)-(c)}.

\noindent To prove our numerical considerations, in-situ diagnostic in form of a pump-probe microscope is used to observe the material's interaction of a single laser pulse in non-strengthened Corning Gorilla$^{\text{\textregistered}}$ glass and to analyze the growing aberrations directly. 

\noindent A single seed pump-probe system with two separate amplifier beamlines for pump and probe is used.
The pulse duration of the infrared processing laser pulses with a wavelength of $\lambda = \unit[1030]{nm}$ can be adjusted with a variable compressor from \unit[400]{fs} up to \unit[20]{ps}. Beam shaping of the initial Gaussian laser pulse occurs, as described in the previous section, at the following SLM and is imaged demagnified into the glass sample, which can be placed with help of a translation axis in the $x$-$y$-plane and adjusted manually for the tilt angle $\theta$, see \hyperref[fig:Figure1]{Fig.\,\ref{fig:Figure1}(a)} and \hyperref[fig:Figure1]{\ref{fig:Figure1}(b)}. 
\noindent To create temporal snapshots of the interaction events, a transmission microscope is implemented transversely to the processing direction. The appropriate illumination is provided by the second amplifier in form of a $\unit[300]{fs}$ laser pulse with a wavelength of $\lambda= \unit[515]{nm}$, see \hyperref[fig:Figure1]{Fig.\,\ref{fig:Figure1}(c)}. An additional bandpass filter with center wavelength of $\unit[514.5]{nm}$ and spectral width of $\pm \unit[1]{nm}$ is used to reduce plasma luminescence and scattered light. The described master oscillator power amplifier (MOPA) setup of the pump-probe system guarantees the synchronization between the two laser pulses and offers the possibility to add an internal time delay. Together with an external delay line for fine-tuning a sub-picosecond resolution over milliseconds can be achieved. 
%Verrechnung Bilder
Single shot experiments are conducted throughout the whole paper. Each experiment is carried out on a pristine position to ensure consistency and therefore reproducibility. We apply the formalism from \cite{Grossmann:16} to calculate the optical depth $\tau(x,z) = \ln\left[I_0(x,z)/I_S(x,z)\right]$. An increased optical depth can result from absorption due to free electrons, but also from scattered light or loss-sources like transient states or voids. 
%Auswertung Experimente

\noindent \hyperref[fig:Figure4]{ Figure\,\ref{fig:Figure4}(d)} shows the measured optical depth for a single Bessel-like laser pulse with a pulse duration of $\unit[1]{ps}$, $\unit[30]{\upmu J}$ pulse energy and $\theta=\unit[0]{^\circ}$. The time delay for all the pump-probe measurements shown is $\unit[100]{ps}$ after first interaction between the material and the laser pulse, where the absorption is completed and no further carrier excitation occurs \cite{Gattass:08}. The image shows an interaction-zone in longitudinal $z$-direction with a total length of more than $\unit[900]{\upmu m}$ and less than $\unit[5]{\upmu m}$ in transverse $x$-direction over the whole pronounced area with a maximum optical depth value of $\tau_\text{max}=1.2$.\\
\noindent To compare the experimental data with the theoretical simulations, we want to discuss the relation between our simulations and in-situ experiments. The wave-optical simulations show the calculated intensity distribution without considering nonlinear effects. The experimental pump-probe images on the other hand show the effect of absorption inside the material including linear and nonlinear effects. The primary absorption to excite electrons from the valence band to the conduction band is nonlinear \cite{Brabec2000}. Then, complex interaction due to different absorption mechanisms occurs until no further energy is supplied, see \cite{Schaffer2001,Stuart1995}. Furthermore, relaxation processes set in. In this context transient defects like self-trapped excitons, non-bridging oxygen-hole centers and electron vacancies have been discussed for fused silica \cite{Mao2004, Chichkov1996}. This becomes more complex for glasses with additional components. The defects show a reduced ionization energy compared to the pristine material. Hence, these regions exhibit an increased contrast in our shadowgraph images due to early absorption of the probe beam. Our observation at $\unit[100]{ps}$ shows the effect resulting from the electron density in the material. Assuming a short diffusion length, defects should appear close to the regions, where the electron density was increased. Comparison of \hyperref[fig:Figure4]{ Fig.\,\ref{fig:Figure4}(a)} and \hyperref[fig:Figure4]{(d)} indicates matching profiles for the linear simulated intensities and the experimental data. Here, the linear simulations are sufficient to describe the interaction of the Bessel-like beams. This is important to note because other studies with different beam profiles, e.g. Gaussian beams \cite{Grossmann:17}, show a visible discrepancy between experiment and theory.

\noindent The measured data in \hyperref[fig:Figure4]{ Fig.\,\ref{fig:Figure4}(e)} and \hyperref[fig:Figure4]{(e*)} show the effect of a tilting angle $\theta=\unit[30]{^\circ}$,  or $\theta'\approx\unit[20]{^\circ}$ according to Snell's law with an assumed refractive index of $n\approx 1.5$ for a wavelength of $\lambda = \unit[1030]{nm}$ in Corning Gorilla$^{\text{\textregistered}}$ glass. The former sharp profile is shortened to less than $\unit[700]{\upmu m}$ and broadened to more than $\unit[20]{\upmu m}$ with the result of optical depth attenuation to $\tau_\text{max}=0.2$. To understand this loss, the normalized cross-section insets in \hyperref[fig:Figure4]{ Fig.\,\ref{fig:Figure4}(a)} and \hyperref[fig:Figure4]{(b)} are helpful. The two insets show the $x$-$y$-plane of the adjacent profile at two different positions (red and green line). Whereas the insets of \hyperref[fig:Figure4]{ Fig.\,\ref{fig:Figure4}(a)} remain confined inside the material, the profile in \hyperref[fig:Figure4]{ Fig.\,\ref{fig:Figure4}(b)} splits up with increasing depth, whereby the peak intensities are weakened. Due to this attenuation the threshold of the aforementioned nonlinear absorption is barely reached, the carrier generation cascade is limited and as a result no reliable cleaving process can be developed with such a profile. More absorption projections for different angles of $\theta$ are pointed out in \href{https://osapublishing.figshare.com/articles/Visualization3_mp4/6081107}{Visualisation\,3}.

%Korrekturdiskussion
\noindent The pump-probe measurement in \hyperref[fig:Figure4]{ Fig.\,\ref{fig:Figure4}(f)} shows the successful implementation of the aberration correction approach described above with a measured resulting angle $\theta'=\unit[19.7]{^\circ}$. Elongation in lateral and confinement in transverse direction, optical depth values, and the intensity distribution are similar to the aforeshown undisturbed beam-profile in \hyperref[fig:Figure4]{ Fig.\,\ref{fig:Figure4}(c)}. However, there is a slight curvature on the corrected beam profile, but the visual impression is deceptive because the $x$-axis is scaled by a factor of 5 and the deviation from the beam center to the edges is only $\approx \unit[3]{\upmu m}$ by a total length of roughly $\unit[1]{mm}$. Again, the pump-probe measurement shows that a linear simulation model is sufficient. 

%Fehlerdiskussion
\noindent The highest achieved chamfer angles of $\theta'\approx 20^\circ$ presented here are initially limited by the employed microscope objective with given NA of $0.4$ of which already up to $90\%$ are used by the far-field ring-distribution of the illuminated digital axicon, cf. \hyperref[fig:Figure1]{Fig\,\ref{fig:Figure1}(c),\,(d)}. Applying the phase correction to this axicon grating even higher spatial frequencies are displayed by the SLM that are no longer accepted by our microscope objective for $\theta > 30^\circ$. This issue could be compensated to a certain extend using a smaller axicon angle $\alpha$. However, in our case, the next limitation represents the objective's housing which collides with the glass substrate for $\theta \gtrsim 40^\circ$. Depending on the thickness of the glass substrate to be processed and the available pulse energy a different microscope objective (with adapted NA and working distance) could lead to higher $\theta$-values. If the tilt angle is further increased ($\theta > 50^\circ$), Fresnel losses at the surface would require an additional amplitude correction.
\begin{figure}[]
\centering
\includegraphics[width=0.90\linewidth]{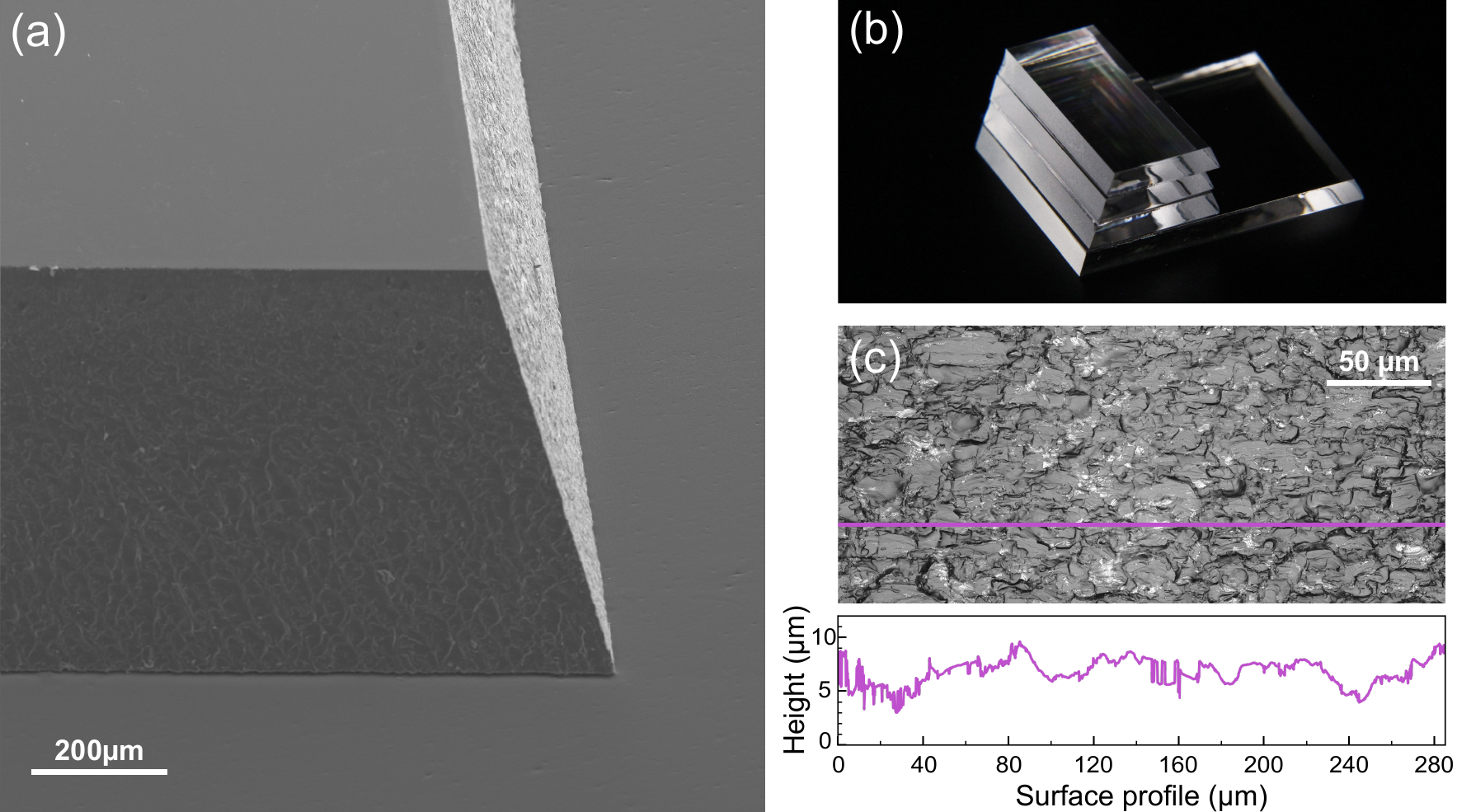}
\caption{Scanning electron microscope image (a) for the cleaved edges with $\unit[30]{^\circ}$ tilting angle in air of $\unit[1]{mm}$ glass and cleaved samples of $\unit[2]{mm}$ thickness in (b), both SCHOTT borofloat 33. Close-up on the cleaved surface and its corresponding surface-roughness measurement in (c).}
\label{fig:Figure5}
\vspace{-0.5cm}
\end{figure}

\noindent To demonstrate the applicability of our approach, we integrated the resurrected features of the corrected Bessel-like beams into a cleaving process. The employed processing laser was changed to a $\unit[120]{W}$ disk-laser in order to process glasses with thicknesses of up to $\unit[2]{mm}$ (SCHOTT borofloat$^{\text{\textregistered}}$ 33). Here we report, to the best of our knowledge, on the first single pass edge cleaving laser application with $\theta'=\unit[20]{^\circ}$ inside the material, cf. \hyperref[fig:Figure5]{ Fig.\,\ref{fig:Figure5}(a),(b)}. We use a spatial pulse distance of $\unit[6]{\upmu m}$ at a processing speed of $\unit[40]{mm/s}$ combined with a pulse train configuration (burst) consisting of four pulses with a temporal delay of $\unit[17]{ns}$ between each consecutive pulses, pulse duration of $\unit[1]{ps}$ and a pulse train energy of approx. $\unit[800]{\upmu J}$. \hyperref[fig:Figure5]{Figure\,\ref{fig:Figure5}(a)} shows a scanning electron microscope (SEM) image of two cleaved glass edges and reveals the precise and defined edge quality for all inclined edges on the $\unit[1]{mm}$ thick glass sample. One exemplary image of the surface quality created with a laser scanning microscope and one corresponding line plot is depicted in \hyperref[fig:Figure5]{Fig.\,\ref{fig:Figure5}(c)}. The arithmetic mean roughness for this process is $R_a \approx \unit[0.8]{\upmu m}$ and absolute vertical height offset $R_q \approx \unit[5]{\upmu m}$, which are comparable to the roughness achieved by accelerating beams \cite{Mathis2012}. \\
%Zusammenfassung
\noindent In conclusion, we presented a detailed wave-optical aberration analysis occurring for beam transition at tilted interfaces and diligent correction thereof made accessible via digital holography. Experimental in-situ pump-probe diagnostics were applied to examine the emerging aberrations and verify the efficacy of the approach. In a final experiment we demonstrated the successful integration of our concept into an existing glass-cleaving system and were able to cut inclined edges for glass thicknesses up to $\unit[2]{mm}$ with an arithmetic mean roughness of sub-micron quality. Future concepts may include combination of different cleaving strategies, paving the way to inner contours or more advanced edge-profiles like chamfers and round edges.

%Beschreibung Schnitt

\section*{Funding Information}

Federal Ministry of Education and Research (BMBF)(13N13927, 13N13930), TRUMPF Laser- und Systemtechnik GmbH.

\vspace{0.3cm}
\noindent\textbf{Disclosures.} The authors declare no conflicts of interest.

\end{document}